\documentclass[a4paper,12pt]{article}

\pdfoutput=1 

\usepackage[hmargin=.7in,vmargin=1.1in]{geometry}
\usepackage{indentfirst}
\usepackage{amsfonts}
\usepackage{mathrsfs}
\usepackage{amsmath}
\usepackage{amssymb}
\usepackage{authblk}
\usepackage{cite}
\usepackage{xcolor}
\usepackage{mathtools}
\usepackage{tensor}
\usepackage{graphicx}
\usepackage{bm}
\usepackage{upgreek}
\usepackage{braket}
\usepackage{color,soul}
\usepackage{csquotes}
\usepackage{caption}
\usepackage{subcaption}
\usepackage{multirow}
\usepackage{lipsum}
\usepackage{xcolor,diagbox}
\usepackage{slashed}
\usepackage{epsfig,makeidx,graphics,latexsym}
\newcommand{\abs}[1]{\left\lvert#1\right\rvert}

\usepackage[bookmarksnumbered=true,bookmarksopen=true]{hyperref}
 \hypersetup{colorlinks,
             linkcolor=[rgb]{0,0.3,0.6}, 
             citecolor=[rgb]{0,0.3,0.6}, 
             urlcolor=[rgb]{0,0.3,0.6}}

\newcommand {\mr}    {\mathrm}
\newcommand {\ii}    {\mathrm{i}}
\newcommand {\ee}    {\mathrm{e}}

\def\dif{\mathop{}\hphantom{\mskip-\thinmuskip}\mathrm{d}}%
\let\daccent\d
\let\d\relax
\newcommand\d{\ifmmode\dif\else\expandafter\daccent\fi}

\begin{document}

\title{\Large\textbf{Perturbations of Einstein--Maxwell--phantom spacetime: Instabilities of charged Ellis--Bronnikov wormholes and quasinormal modes of black holes}}

\author{Guan-Yu Wu\thanks{wugy@mail.nankai.edu.cn}}

\author{Si-Yu Wang\thanks{wangsiyu@mail.nankai.edu.cn}}

\author{Yan-Gang Miao\thanks{Corresponding author: miaoyg@nankai.edu.cn.}}

\affil{\normalsize{\em School of Physics, Nankai University, 94 Weijin Road, Tianjin 300071, China}}

\date{}

\maketitle

\begin{abstract}
Phantom scalar fields, as a viable candidate for dark energy, have been instrumental in eliminating spacetime singularities and constructing wormholes and regular black holes. We investigate the Einstein--Maxwell--phantom (EMP) framework, in which the Ellis--Bronnikov wormholes can be charged and regular black holes can be admitted. While the previous study has shown the stability of EMP wormholes under massless scalar field perturbations, we further perform a comprehensive linear analysis of the EMP spacetime through gravito-electromagnetic field perturbations in the axial sector and phantom scalar field perturbations under an approximate treatment  in the polar sector. Our analyses of effective potentials and finite difference time profiles reveal the linear instability of EMP wormholes. In the black hole scenario, the quasinormal spectra of Type I black holes, where the matrix-valued direct integration method and the Prony method are used, recover those of general relativity (GR) when the scalar charge goes to zero. Finally, by introducing the concepts of generalized specific charge and mixing angle, we quantify how the relative contributions between the phantom scalar and the electromagnetic fields modify the quasinormal spectra, and we assess the prospects for detecting spectral deviations between the EMP theory and GR in gravitational wave observation.
\end{abstract}

\tableofcontents

\section{Introduction}
\label{sec:intr}

Ever since the spectroscopic and photometric observations of Type Ia supernovae in the late 20th century, the cosmological data have revealed~\cite{SupernovaSearchTeam:1998fmf,SupernovaCosmologyProject:1998vns,SupernovaSearchTeam:2004lze,SNLS:2005qlf,SupernovaCosmologyProject:2011ycw} that the universe is undergoing accelerated expansion. The acceleration is typically attributed to dark energy, whose equation of state is defined by the ratio of its pressure $p$ and energy density $\rho$, i.e.,
\begin{align}
    w\equiv p/\rho\label{EoS},
\end{align}
with $w<-1/3$ required for cosmic acceleration. Within the widely accepted $\Lambda$CDM model, dark energy is realized as the cosmological constant $\Lambda$, for which $w=-1$~\cite{Carroll:2000fy}. Alternative scenarios, such as quintessence models~\cite{Wetterich:1987fm, Zlatev:1998tr} and K-essence models~\cite{Armendariz-Picon:2000nqq, Armendariz-Picon:2000ulo}, allow $w>-1$, whereas phantom dark energy $(w<-1)$ would lead to a future Rip~\cite{Caldwell:1999ew,Sami:2003xv,Nojiri:2005sx,Frampton:2011sp,Frampton:2011aa}. Although the current observations show remarkable consistency with $\Lambda$CDM, they do not exclude a slightly phantom behavior: Several independent probes suggest $w\sim-1.03$~\cite{Planck:2018vyg,DES:2022ccp,Giare:2025pzu}, and a combined analysis yields $w=-1.013^{+0.038}_{-0.043}$~\cite{Escamilla:2023oce}.

In Einstein's general theory of relativity, the acceleration of the universe can be accounted for either by adding a positive cosmological constant $\Lambda$ on the left-hand side of the Einstein equation or by introducing exotic matter with negative pressure to the source term on the right-hand side~\cite{Sami:2003xv}. The former prescription realizes Gliner's $\mu$-vacuum~\cite{Gliner:1966} in a de Sitter geometry, which can prevent spacetime singularity arising in gravitational collapse. A spacetime singularity, where the matter density and spacetime curvature diverge, marks the breakdown of physics. According to the Penrose singularity theorem~\cite{Penrose:1965}, a singularity is inevitable within the framework of general relativity and under reasonable energy conditions, but the cosmic censorship conjecture states~\cite{Penrose:2002} that such singularities are always hidden behind event horizons, i.e., no naked singularities exist. If the latter prescription is adopted, source terms that violate certain energy conditions are indispensable in order to construct regular (nonsingular) black holes. Notable examples include magnetically charged solutions in nonlinear electrodynamics~\cite{Ayon-Beato:2000mjt,Bronnikov:2000vy,Fan:2016hvf} and configurations with phantom scalar hairs~\cite{Bronnikov:2005gm,Babichev:2020qpr,Chew:2022enh,Barrientos:2022avi,Ding:2024qrf} (see Ref.~\cite{Lan:2023cvz} for a brief review).

The existence of black holes has been empirically confirmed through the historical direct detection of gravitational waves from a binary black hole merger by Laser Interferometer Gravitational-Wave Observatory (LIGO) and Virgo~\cite{LIGOScientific:2016aoc}, with numerous subsequent events reinforcing this statement~\cite{LIGOScientific:2016sjg,LIGOScientific:2017bnn,LIGOScientific:2017ycc,KAGRA:2021vkt,LIGOScientific:2025slb}. Moreover, Fermi Gamma-ray Burst Monitor (Fermi-GBM) and International Gamma-Ray Astrophysics Laboratory (INTEGRAL) observed~\cite{LIGOScientific:2017ync, LIGOScientific:2017vwq} a short gamma-ray burst from the same event in 1.74 seconds after LIGO and Virgo detected gravitational waves from the binary neutron star merger --- GW170817, inaugurating an era of multi-messenger astronomy. Furthermore, the first images of the supermassive black hole at the center of M87~\cite{EventHorizonTelescope:2019dse,EventHorizonTelescope:2019uob,EventHorizonTelescope:2019jan,EventHorizonTelescope:2019ths,EventHorizonTelescope:2019pgp,EventHorizonTelescope:2019ggy} and Sagittarius A* in the Milky Way~\cite{EventHorizonTelescope:2022wkp,EventHorizonTelescope:2022apq,EventHorizonTelescope:2022wok,EventHorizonTelescope:2022exc,EventHorizonTelescope:2022urf,EventHorizonTelescope:2022xqj}, captured by Event Horizon Telescope (EHT), provide direct high-resolution evidence for black holes in galactic nuclei. These breakthroughs provide us with more possibilities to test fundamental physical problems, e.g., the singularity problem~\cite{Berti:2015,Cardoso:2016} and the validity of the no-hair theorem~\cite{Berti:2005ys,Berti:2007zu,Isi:2019aib}.

Black holes, as nearly out-of-equilibrium dissipative systems~\cite{Maggio:2020jml}, resonate in a series of quasinormal modes (QNMs) when perturbed or formed after coalescence, a stage also known as ringdown~\cite{Nollert:1998,Berti:2007}. Theoretical study and experimental observations of black hole ringdown is crucial for probing gravity in the strong field regime, testing general relativity and constraining modified gravity theories~\cite{Echeverria:1989hg,Cardoso:2016,Cardoso:2017cqb,Cardoso:2019rvt}. Tentative echoes observed in LIGO/Virgo data from binary black hole mergers have been reported\cite{Abedi:2016hgu}, sparking widespread discussion~\cite{Westerweck:2017hus,Abedi:2018pst,Abedi:2018npz,Abedi:2021tti,Uchikata:2023zcu}. Echo proposals often invoke quantum modifications near an event horizon~\cite{Cardoso:2019apo,Oshita:2019sat,Wang:2019rcf,Almheiri:2012rt,Giddings:2015uzr,Oshita:2018fqu,Chakravarti:2021jbv}, driven by the fact that classical black holes, which lack a natural reflection mechanism, do not inherently produce echo signals, while wormholes naturally generate echoes during the ringdown stage~\cite{Bueno:2017hyj,Bronnikov:2019sbx,Liu:2020qia,Ou:2021efv,Yang:2024prm}.

As with regular black holes, wormhole models also require exotic matter that violates energy conditions, specifically the null energy condition (NEC)~\cite{Morris:1988tu,visser1995lorentzian}. The first eponymous wormhole solution --- now known as the Ellis--Bronnikov (EB) wormhole --- was developed independently by Ellis~\cite{Ellis:1973yv} and Bronnikov~\cite{Bronnikov:1973f} in a theory where the gravity is coupled to a massless phantom scalar field. In general, a phantom field is characterized by a flapping sign in the kinetic term of its action~\cite{Goulart:2017iko}. More recently, a charged EB wormhole was proposed~\cite{Huang:2019arj} by the way to extend a class of Einstein--Maxwell--dilaton (EMD) theory~\cite{Huang:2019lsl} to an Einstein--Maxwell--phantom (EMP) theory through a complex transformation. In this EMP spacetime, sufficient electric charge yields a wormhole geometry, while below a critical charge the solution becomes a regular black hole, which can be classified as Type I or Type II based on its horizon structure. Further study~\cite{Ou:2021efv} indicates that, under massless scalar field perturbations, EMP wormholes remain stable and exhibit echo signals near the wormhole--black hole transition threshold, whereas instabilities of other phantom wormholes have been reported in Refs.~\cite{Gonzalez:2008wd,Gonzalez:2008xk,Gonzalez:2009hn,Bronnikov:2011if,Bronnikov:2012ch}. Note that the EMP spacetime is sourced by the phantom scalar field and the Maxwell field, and the latter can also be phantom-like in certain region. Such phantom-like behaviors imply negative kinetic energy of the source terms and can potentially induce spacetime instabilities. Investigating both phantom scalar and gravito-electromagnetic field perturbations is therefore essential for assessing the dynamical stability of the EMP spacetime and may offer fresh insights into gravity. 

The organization of this paper is as follows. In Sec.~\ref{sec:EMP}, we briefly review the derivation and key properties of the EMP spacetime. In Sec.~\ref{sec:perturbation}, we study the linear perturbations of the EMP spacetime and derive the master equations of gravito-electromagnetic field perturbations in the axial sector and phantom scalar field perturbations in the polar sector, where the latter is discussed within a simplified and decoupled framework. In Sec.~\ref{sec:instability}, we reveal instability of EMP wormholes by exhibiting the singular behavior in the gravito-electromagnetic effective potential and growth of the phantom scalar field perturbation. In Sec.~\ref{sec:numerical}, we compute the quasinormal frequencies (QNFs) of Type I black holes using the matrix-valued direct integration method and the Prony method, and we assess the observability of deviations between the gravitational QNMs of Type I black holes and Reissner--Nordstr{\"o}m (RN) black holes. Finally, the conclusion is made in Sec.~\ref{sec:conclusion}.
\section{Einstein--Maxwell--phantom spacetime}
\label{sec:EMP}

The EMP theory can be obtained from the EMD theory, where the latter arises~\cite{Garfinkle:1990qj} as a four-dimensional low-energy effective action of string theory,
\begin{align}
    S_\mr{EMD}=\int\dif^4 x\sqrt{-g}\left(R-2\left(\partial\phi_\mr{d}\right)^{2}-\frac{1}{4}\ee^{-2\phi_\mr{d}}F^2\right),\label{action_emd}
\end{align}
where $g$ is the metric determinant, $R$ is the Ricci scalar and the dilaton scalar field $\phi_\mathrm{d}$ is minimally coupled to gravity, but nonminimally coupled to the Maxwell field $F=\dif A$ in the form $\ee^{-2\phi_\mathrm{d}}$. Rescaling the dilaton $\phi=2\phi_\mathrm{d}$ yields the standard coupling $\ee^{-\phi}$. Moreover, such a nonminimal coupling is further generalized~\cite{Huang:2019lsl}  to $\left(\gamma_1\cosh{\phi}+\gamma_2\sinh{\phi}\right)^{-1}$, where $\gamma_1$ and $\gamma_2$ are coupling constants, and the generalized coupling function reduces to $\ee^{-\phi}$ when $\gamma_1=\gamma_2=1$. A general classification of phantom EMD black holes has been presented in Ref.~\cite{Clement:2009}, where various causal structures and horizon configurations have been analyzed. The EMP theory considered here can be regarded as a particular extension of those phantom EMD models through the complex transformation,
\begin{align}
    \left(\phi,\gamma_2\right)\to\left(\ii\phi,\ii\gamma_2\right),\label{comp_trans}
\end{align}
one then arrives~\cite{Huang:2019arj} at the EMP action,
\begin{align}
    S=\int\dif^4 x\sqrt{-g}\left(R+\frac{1}{2}(\partial\phi)^{2}-\frac{1}{4}Z^{-1}F^2\right),\qquad Z=\gamma_{1}\cos\phi+\gamma_{2}\sin\phi,\label{EMP_S}
\end{align}
where the scalar field is phantom-like due to its flapping sign of the kinetic term. In addition, since the coupling function $Z$ is not positive definite regardless the value of $\gamma_1$ and $\gamma_2$, the Maxwell field acquires a phantom-like behavior in negative regions of $Z$. The equations of motion associated with the variation of the phantom scalar field $\phi$, the Maxwell field $A_\mu$ and the gravitational field $g_{\mu\nu}$ are given respectively by
\begin{align}
    &\square\phi=-\frac{1}{4}\frac{\partial Z^{-1}}{\partial\phi}F^2,\label{eom_s} \\
    &\nabla_{\mu}(Z^{-1}F^{\mu\nu})=0,\label{eom_em}\\
    &R_{\mu\nu}-\frac{1}{2}Rg_{\mu\nu}=T_{\mu\nu}^{A}+T_{\mu\nu}^{\phi}\label{eom_g},
\end{align}
where the energy--momentum tensors of the Maxwell field $T_{\mu\nu}^{A}$ and the phantom scalar field $T_{\mu\nu}^\phi$ take the forms,
\begin{align}
    T_{\mu\nu}^{A} =\frac{1}{2}Z^{-1}\left(F_{\mu\nu}^{2}-\frac{1}{4}g_{\mu\nu}F^2\right),\qquad T_{\mu\nu}^\phi =-\frac12\partial_{\mu}\phi\partial_{\nu}\phi+\frac14g_{\mu\nu}(\partial\phi)^2.
\end{align}

Assuming a static and spherically symmetric wormhole metric,
\begin{align}
    \dif s^2&=-h(r)\dif t^2+\frac{1}{h(r)}\dif r^2+(r^2+q^2)\dif\Omega^2_2,\label{EMP_ansatz}
\end{align}
one gives~\cite{Huang:2019arj} the charged EB wormhole solution of the EMP theory,
\begin{align}
    \phi=2\arccos\left(\frac{r}{\sqrt{r^2+q^2}}\right),\qquad F_{tr}=-\frac{QZ}{r^2+q^2},\qquad h(r)=1-\frac{\gamma_{2}Q^{2}r}{4q(r^{2}+q^{2})}+\frac{\gamma_{1}Q^{2}}{4(r^{2}+q^{2})}\label{EMP_metric_coef},
\end{align}
where the regularization parameter $q$ corresponds to the radius of the wormhole throat $Q$ is an integration constant that contributes to both the mass and electric charge of the EMP spacetime. Note that the EMP spacetime is asymptotically flat and regular, featuring a wormhole throat (also called a bounce) at $r=0$. The mass $M$ and electric charge $Q_\mathrm{e}$ are
\begin{align}
    M=\frac{\gamma_2 Q^2}{8q},\qquad Q_\mathrm{e}=\gamma_1 Q.\label{mass}
\end{align}

If the following condition is chosen,
\begin{align}
    \left(\frac{Q_\mathrm{e}}{\gamma_1}\right)^2>\frac{8M^2\left(-\gamma_1+\sqrt{\gamma_1^2+\gamma_2^2}\right)}{\gamma_2^2},\label{wormhole_condition}
\end{align}
no real zeros of $h(r)$ exist and the geometry describes a wormhole; otherwise, if
\begin{align}
    \left(\frac{Q_\mathrm{e}}{\gamma_1}\right)^2\leqslant\frac{8M^2\left(-\gamma_1+\sqrt{\gamma_1^2+\gamma_2^2}\right)}{\gamma_2^2},\label{black_hole_condition}
\end{align}
with the equality marking the extremal limit, the real zeros $r_\pm$ appear, corresponding to the horizons of 
a regular black hole,
\begin{align}
    r_{\pm}=\frac{\gamma_2^2 Q^2\pm \sqrt{\gamma_2^2 Q^4-16\gamma_1 q^2Q^2-64q^4}}{8q}.
\end{align}
The horizon distribution of EMP black holes admits two distinct configurations: 
\begin{itemize}
  \item[] \textbf{Type I black hole.} An outer event horizon and an inner Cauchy horizon lie on the same side of the wormhole throat (bounce); consequently the exterior causal structure closely resembles that of RN black holes.
  \item[] \textbf{Type II black hole.} Two horizons straddle the throat, so that the bounce smoothly interpolates between black hole and white hole regions; the global diagram therefore differs qualitatively from the RN scenario.
\end{itemize}
Further details and the Penrose diagram can be found in Ref.~\cite{Huang:2019arj}.

We note that $q$ plays a central role in the EMP theory, functioning not only as a regularization parameter for constructing wormholes and resolving spacetime singularity, but also as a geometric parameter determining the radius of the wormhole throat. It is further interpreted as a scalar charge, in analogy to electric charge. An interesting question: Is there a physically motivated range of $q$, for instance, from the energy conditions? The answer is negative. The phantom scalar field that sources the EMP solutions explicitly violates the null energy condition (NEC)~~\cite{Morris:1988tu,visser1995lorentzian}, and such a condition is a prerequisite for constructing traversable wormholes. Moreover, the NEC is the weakest of the four energy conditions. That is,  if the NEC is violated, the other three energy conditions, the weak, strong, and dominant energy conditions, are also violated. Therefore, the energy conditions cannot be used to impose meaningful constraints on the scalar charge $q$.
On the other hand, the EMP framework, as a result of the four-dimensional low-energy effective theory of string theory,  must recover general relativity in the classical limit. This is achieved under the parameter choice,
\begin{align}
    \gamma_1=\frac{1}{4},\qquad\gamma_2=\frac{8qM}{Q^2},\qquad q\to 0^+,\label{limit_to_GR}
\end{align}
which implies that the quantum-gravity consistency sets a lower bound for $q$, namely zero. This limit is consistent with fundamental principles of general relativity, e.g., the preservation of U(1) gauge symmetry, enforcement of the no-scalar-hair theorem, and satisfaction of the positive mass theorem~\cite{Schon:1979rg}, thereby establishing theoretical consistency of the EMP theory with the Einstein-Maxwell dynamics.

\section{Perturbation equations}
\label{sec:perturbation}

\subsection{General set up}
\label{sec:setup}
In this section, we study the linear perturbation of the EMP spacetime. The phantom scalar field, Maxwell field and gravitational field as perturbations can be written as
\begin{align}
    \phi&\to\bar{\phi}+\delta\phi,\label{3.2}\\
    A_{\mu}&\to \bar{A}_{\mu}+\delta  A_{\mu}\label{3.3},\\
    g_{\mu\nu}&\to \bar{g}_{\mu\nu}+\delta  g_{\mu\nu},
\end{align}
where the quantities with a bar represent background fields, and those with a $\delta$ denote the perturbation fields. Following Chandrasekhar's procedure~\cite{Chandrasekhar}, the perturbed spacetime is characterized by a general non-stationary axisymmetric line element,
\begin{align}
    \dif s{^{2}}=-\ee{^{2\nu}}\dif t{^{2}}+\ee{^{2\psi}}\left(\dif\varphi-q{_{2}}\dif x^2-q{_{3}}\dif x^3-\sigma\dif t\right)^{2}+\ee{^{2\mu_{2}}}(\dif x^2)^2+\ee{^{2\mu_{3}}}(\dif x^3)^2,\label{general_spacetime}
\end{align}
where $q_2$, $q_3$, $\sigma$, $\nu$, $\psi$, $\mu_2$ and $\mu_3$ are functions of spacetime coordinates ($t, r, \theta$) and admit the freedom of gauge~\cite{Chandrasekhar},
\begin{align}
    (\sigma_{,2}-q_{2,0})_{,3}-(\sigma_{,3}-q_{3,0})_{,2}+(q_{2,3}-q_{3,2})_{,0}=0,\label{guage-freedom}
\end{align}
where the comma denotes a partial derivative with respect to the local coordinates.

The charged EB wormhole solution of the EMP theory~\cite{Huang:2019arj}, as a special case of Eq.~(\ref{general_spacetime}), has the following coefficients,
\begin{align}
    e^{2\nu}=e^{-2\mu_{2}}=h(r)=\frac{\Delta}{\varrho^{2}},\quad e^{2\psi}=\varrho^{2}\sin^{2}{\theta},\quad e^{2\mu_{3}}=\varrho^{2},\quad
    q_{2}=q_{3}=\sigma=0,\quad \varrho^{2}=r^{2}+q^{2}.\label{coefficients_emp}
\end{align}
Consequently, a general perturbation of the EMP spacetime gives rise to non-vanishing $q_2$, $q_3$ and $\sigma$, which are small quantities of the linear order that introduce a rotation to the spacetime and are referred to as axial (odd-parity) gravitational field perturbations. Meanwhile, the functions $\nu$, $\mu_2$, $\mu_3$ and $\psi$ acquire small increments $\delta\nu$, $\delta\mu_2$, $\delta\mu_3$ and $\delta\psi$, which do not induce any rotation and are known as polar (even-parity) gravitational field perturbations.

In accordance with Chandrasekhar's procedure~\cite{Chandrasekhar}, we also implement the tetrad formalism using the basis covectors,
\begin{alignat}{5}
e_{0\mu} &{}={}& \bigl(&\ \ -\ee^\nu,       &0,           &\quad\qquad0,              &\quad 0 \bigr),\notag\\[1mm]
e_{1\mu} &{}={}& \bigl(&-\sigma \ee^\psi,   &\ee^\psi,    &\ -q_{2}\ee^\psi,          &\ -q_{3}\ee^\psi \bigr),\notag\\[1mm]
e_{2\mu} &{}={}& \bigl(&\ \qquad0,          &\ \qquad0,   &\,\qquad \ee^{\mu_{2}},    &\quad 0 \bigr),\notag\\[1mm]
e_{3\mu} &{}={}& \bigl(&\ \qquad0,          &0,           &\quad\qquad0,              &\quad \ee^{\mu_{3}} \bigr)\label{tetrad}.
\end{alignat}
The corresponding basis vectors can be obtained by contracting the basis covectors in Eq.~(\ref{tetrad}) with the inverse of metric  Eq.~(\ref{general_spacetime}). Such a formalism determines a local inertial frame with the Minkowski metric, $\eta_{ab}={e_a}^{\mu}e_{b\mu}$.

To continue discussions for simplicity, we set
\begin{align}
    P_{ab} ={e_{a}}^{\mu}{e_{b}}^{\nu}Z^{-1}F_{\mu\nu},
\end{align}
so that the equation of motion Eq.~(\ref{eom_em}) takes the standard form,
\begin{align}
    \nabla^a P_{ab}=0\label{EMp},
\end{align}
which consists of four component equations,
\begin{align}
    (\ee^{\psi+\mu_{3}}P_{02})_{,2}+(\ee^{\psi+\mu_{2}}P_{03})_{,3}&=0,\notag\\
    (\ee^{\psi+\mu_{2}}P_{03})_{,0}-(\ee^{\psi+\nu}P_{23})_{,2}&=0,\label{maxwell_1}\\
    (\ee^{\psi+\mu_{3}}P_{02})_{,0}+(\ee^{\psi+\nu}P_{23})_{,3}&=0,\notag\\
    (\ee^{\mu_{2}+\mu_{3}}P_{01})_{,0}+(\ee^{\nu+\mu_{3}}P_{12})_{,2}+(\ee^{\nu+\mu_{2}}P_{13})_{,3}&=\ee^{\psi+\mu_{3}}P_{02}Q_{02}+\ee^{\psi+\mu_{2}}P_{03}Q_{03}-\ee^{\psi+\nu}P_{23}Q_{23}\notag,
\end{align}
where $Q_{AB}\equiv q_{A,B}-q_{B,A}$ and $Q_{A0}\equiv q_{A,0}-\sigma_{,A}$ with $A,B=2,3$. Although $P_{ab}$ does not satisfy the Bianchi identity and is thus not the true Maxwell field, we refer to $P_{ab}$ as the electromagnetic field in our perturbation analysis and reserve the term Maxwell field exclusively for the genuine field $F_{ab}$.

The Bianchi identity for the Maxwell field,
\begin{align}
    \nabla_{[c}F_{ab]}=0,
\end{align}
also has four component equations,
\begin{align}
    (\ee^{\psi+\mu_{2}}F_{12})_{,3}+(\ee^{\psi+\mu_{3}}F_{31})_{,2}&=0,\notag \\
    (\ee^{\psi+\nu}F_{01})_{,2}+(\ee^{\psi+\mu_{2}}F_{12})_{,0}&=0,\label{maxwell_2}\\
    (\ee^{\psi+\nu}F_{01})_{,3}+(\ee^{\psi+\mu_{3}}F_{13})_{,0}&=0,\notag\\
    (\ee^{\nu+\mu_{2}}F_{02})_{,3}-(\ee^{\nu+\mu_{3}}F_{03})_{,2}+(\ee^{\mu_{2}+\mu_{3}}F_{23})_{,0}&=\ee^{\psi+\nu}F_{01}Q_{23}+\ee^{\psi+\mu_{2}}F_{12}Q_{03}-\ee^{\psi+\mu_{3}}F_{13}Q_{02}\notag.
\end{align}

Eqs.~(\ref{maxwell_1}) and~(\ref{maxwell_2}) can be reorganized into two sets of four equations, one for the axial sector and the other for the polar sector, each containing only three independent degrees of freedom.

\subsection{Gravito-Electromagnetic field perturbation}
For gravito-electromagnetic field perturbations, the linearization of the Einstein equation Eq.~(\ref{eom_g}) yields
\begin{align}
    \delta R_{ab}&=\frac{1}{2}\biggl(F_{ab}^{2}-\frac{1}{4}\eta_{ab}F^{2}\biggr)\frac{\partial Z^{-1}}{\partial\phi}\delta\phi+\frac{1}{2}Z^{-1}\biggl(\delta F_{am}{F_{b}}^{m}+F_{am}\delta {F_{b}}^{m}-\frac{1}{2}\eta_{ab}F^{mn}\delta F_{mn}\biggr) \nonumber\\ 
    &\quad-\frac{1}{2}\delta\left(\partial_{a}\phi\partial_{b}\phi\right)\label{linearized_gravitation_eom}.
\end{align}
The left-hand side of the above equation can be computed via Cartan's equations of structure applied to the 1-forms in Eq.~(\ref{tetrad}). In particular, the axial components $\delta R_{12}$ and $\delta R_{13}$ take the form,
\begin{align}
    &(\sigma_{,2}-q_{2,0})_{,0}=-\frac{1}{\varrho^{4}\sin^{3}{\theta}}\frac{\partial\mathcal{Q}}{\partial\theta}+\frac{QZe^\nu}{\varrho^{3}\sin^{2}{\theta}}B,\label{1R12}\\
    &(\sigma_{,3}-q_{3,0})_{,0}=\frac{\Delta}{\varrho^{4}\sin^{3}{\theta}}\frac{\partial\mathcal{Q}}{\partial r},\label{1R13}
\end{align}
where $\mathcal{Q}(r,\theta,t)=\varrho^{2}\ee^{2\nu}Q_{23}\sin^{3}\theta$ encodes gravitational field perturbations and $B(r,\theta)=\delta P_{01}\sin{\theta}$ represents electromagnetic field perturbations. Note that those terms involving $\delta\left(\partial_{a}\phi\partial_{b}\phi\right)$ vanish in Eqs.~(\ref{1R12}) and~(\ref{1R13}) by virtue of the axial symmetry of the background metric Eq.~(\ref{general_spacetime}). Additionally, the phantom scalar (spin-0) perturbations are parity-even and therefore appear only in the polar (even-parity) sector in the static and spherically symmetric EMP spacetime; they do not source the axial (odd-parity) equations.

Returning to the axial sector and assuming a time dependence $\ee^{-\ii\omega t}$ in the perturbations, elimination of $\sigma$ in Eqs.~(\ref{1R12}) and~(\ref{1R13}) yields the first master equation for gravito-electromagnetic field perturbations,
\begin{align}
    \varrho^{4}\frac{\partial}{\partial r}\left(\frac{\Delta}{\varrho^{4}}\frac{\partial\mathcal{Q}}{\partial r}\right)+\sin^{3}{\theta}\frac{\partial}{\partial\theta}\left(\frac{1}{\sin^{3}{\theta}}\frac{\partial\mathcal{Q}}{\partial\theta}\right)+\omega^{2}\frac{\varrho^{4}}{\Delta}\mathcal{Q}=QZ\ee^\nu\varrho\sin^{3}{\theta}\frac{\partial}{\partial\theta}\left(\frac{B}{\sin^{2}{\theta}}\right).\label{main1}
\end{align}

As mentioned earlier, the axial electromagnetic field equations have three independent components, so as to their linearized forms,
\begin{align}
    &(\ee^{\psi+\nu}\delta F_{01})_{,2}+(\ee^{\psi+\mu_{2}}\delta F_{12})_{,0}=0,\notag \\
    &(\ee^{\psi+\nu}\delta F_{01})_{,3}+(\ee^{\psi+\mu_{3}}\delta F_{13})_{,0}=0, \\
    &(\ee^{\mu_{2}+\mu_{3}}\delta P_{01})_{,0}+(\ee^{\nu+\mu_{3}}\delta P_{12})_{,2}+(\ee^{\nu+\mu_{2}}\delta P_{13})_{,3}=\ee^{\psi+\mu_{3}}P_{02}Q_{02}.\notag
\end{align}
Combining the above three equations and eliminating $\sigma$ yields the second master equation for gravito-electromagnetic field perturbations,
\begin{align}
    \left[Z^{-1}\ee^{2\nu}\left(Z\varrho \ee^{\nu}B\right)_{,2}\right]_{,2}+\frac{\ee^{\nu}}{\varrho}\left(\frac{B_{,3}}{\sin{\theta}}\right)_{,3}\sin{\theta}+\left(\omega^{2}\varrho\ee^{-\nu}-\frac{Q^{2}}{\varrho^{3}}Z\ee^{\nu}\right)B=-Q\frac{\mathcal{Q}_{,3}}{\varrho^{4}\sin{\theta}}.\label{main2}
\end{align}
The variables $r$ and $\theta$ in Eqs.~(\ref{main1}) and~(\ref{main2}) can be separated by the substitutions,
\begin{align}
    &\mathcal{Q}(r,\theta)=\mathcal{Q}(r)C_{l+2}^{-3/2}(\theta),\\
    &B(r,\theta)=\frac{B(r)}{\sin\theta}\frac{\dif C_{l+2}^{-3/2}(\theta)}{\dif\theta}=3B(r)C_{l+1}^{-1/2}(\theta),
\end{align}
where the Gegenbauer function $C_{\nu}^{n}(\theta)$ satisfies
\begin{align}
    \left[\frac{\dif}{\dif\theta}\sin^{2\nu}\theta\frac{\dif}{\dif\theta}+n(n+2\nu)\sin^{2\nu}\theta\right]C_n^\nu(\theta)=0,
\end{align}
and has the recurrence relation,
\begin{align}
    \frac{1}{\sin\theta}\frac{\dif C_n^\nu}{\dif\theta}=-2\gamma C_{n-1}^{\nu+1}.
\end{align}

After separating variables, the gravito-electromagnetic field perturbation is governed by the coupled master equations,
\begin{align}
        &\Delta\frac{\dif}{\dif r}\left(\frac{\Delta}{\varrho^{4}}\frac{\dif}{\dif r}\mathcal{Q}(r)\right)-\mu^{2}\frac{\Delta}{\varrho^{4}}\mathcal{Q}(r)+\omega^{2}\mathcal{Q}(r)=-\frac{QZ\mu^{2}}{\varrho^{3}}\Delta\ee^\nu B(r),\label{electromagnetic master euqation}\\
        &\frac{\dif}{\dif r}\left[Z^{-1}\ee^{2\nu}\frac{\dif}{\dif r}\left(Z\varrho\ee^{\nu}B(r)\right)\right]-(\mu^{2}+2)\frac{\ee^{\nu}}{\varrho}B(r)+\left(\omega^{2}\varrho\ee^{-\nu}-\frac{Q^{2}}{\varrho^{3}}Z\ee^{\nu}\right)B(r)=-\frac{Q}{\varrho^{4}}\mathcal{Q}(r),\label{gravito master euqation}
\end{align}
where $\mu^2=(l-1)(l+2)$. To eliminate the first-order derivative, we introduce two functions $H_1$ and $H_2$,
\begin{align}
\sqrt{Z}\varrho \ee^{\nu}B(r)=-\frac{H_1}{\mu}    ,\qquad \mathcal{Q}(r)=\varrho H_2,
\end{align}
so that $H_1$ and $H_2$ represent electromagnetic and gravitational field perturbations, respectively, with subscripts denoting the spin of the fields. We further adopt the tortoise coordinate $r_*$ defined by
\begin{align}
    \dif r_*=\frac{1}{h(r)}\dif r,\label{tortoise}
\end{align}
and then change the master equation to the Schr{\"o}dinger-like form,
\begin{align}
    \left(\frac{\dif^2}{\dif r_*^2}+\omega^2-\mathbf V\right)\mathbf{H} =0,\label{master_eq}
\end{align}
where
\begin{align}
    \mathbf{H}=\left(\begin{array}{c}H_1\\H_2\end{array}\right),\qquad \mathbf V= \left(\begin{array}{cc}V_{11}&V_{12}\\V_{21}&V_{22}\end{array}\right),
\end{align}
and
\begin{align}
    V_{11}&=\frac{3}{4Z^{2}}\left(\frac{\dif Z}{\dif r_{*}}\right)^{2}-\frac{1}{2Z}\frac{\dif^{2} Z}{\dif r_{*}^{2}}+\left(\mu^{2}+2\right)\frac{\Delta}{\varrho^{4}}+\frac{ZQ^{2}\Delta}{\varrho^{6}},\nonumber \\
    V_{12}&=V_{21}=\frac{Q\sqrt{Z}\mu\Delta}{\varrho^{5}},\nonumber \\ V_{22}&=\mu^{2}\frac{\Delta}{\varrho^{4}}+\frac{2\Delta^2 r^2}{\varrho^{8}}-\frac{\Delta}{\varrho^3}\frac{\dif}{\dif r}\left(\frac{\Delta r}{\varrho^{3}}\right).\label{V GEM}
\end{align}

In the RN background, the effective potential matrix in the master equation can be diagonalized by a constant matrix, yielding decoupled perturbation equations. However,  the coupling cannot be removed in this manner\footnote{It has come to our attention that some studies~\cite{Fernando:2004pc,Lee:2020iau,Wu:2022eiv} have inappropriately applied the RN decoupling procedure in their work. To avoid further misleading, we explicitly clarify here that the RN decoupling procedure is not valid in the EMP spacetime.} in the EMP spacetime, so we shall apply numerical methods directly to the coupled system.

While the primary focus of this work is on the axial perturbation of the EMP spacetime, a brief discussion of the polar sector is warranted to provide a broader context for the stability analysis. A complete characterization of the linear dynamics requires the investigation of both sectors. Below we present a concise qualitative discussion of polar perturbations. We also include a approximate quantitative treatment to provide a preliminary outlook on the polar sector.
\subsection{Phantom scalar field perturbation}
In a spherically symmetric background, perturbations can be decomposed to the sectors with a definite parity. The axial sector (odd-parity), which we have studied in detail, is associated with frame dragging and with coupled gravito-electromagnetic degrees of freedom. By contrast, the polar sector (even-parity) contains a richer set of dynamics, including radial pulsations of the spacetime geometry, variations in the matter distribution (energy--momentum tensor), and the even-parity components of the Maxwell field.

Within the EMP theory, the polar perturbation is expected to be considerably more complex than its axial counterpart. This increased complexity stems from the fact that all background fields --- the EMP metric field $g_{\mu\nu}$, the Maxwell field $A_\mu$, and the phantom scalar field $\phi$ --- are of polar type; accordingly, their linear perturbations will inherently couple to one another, leading to a gravito-electromagnetic-phantom scalar coupled system with dynamics distinct from that of the axial sector. In particular:

\begin{itemize}
    \item In addition to the gravito-electromagnetic degrees of freedom, the polar spectrum contains extra modes sourced by perturbations of the phantom scalar field. The phantom scalar field perturbation is influenced only by the perturbed electromagnetic field. By contrast, the electromagnetic field perturbations are influenced by both the gravitational and phantom scalar field perturbations, and similarly, the gravitational perturbations are influenced by both the electromagnetic and phantom scalar field perturbations, leading to a fully coupled system.
    \item The effective potentials in the polar sector may have qualitatively different structures, which can change stability thresholds for wormholes and shift quasinormal frequencies relative to the axial sector.
    \item The breaking of isospectrality between the axial and polar modes has been reported in other theories with scalar hairs (see, e.g.,~Ref. \cite{Berti:2003ud,Wagle:2021tam}), because the scalar field enters the polar equations but not the axial ones; similar isospectral-breaking is therefore foreseeable in the EMP spacetime.
\end{itemize}

A comprehensive analysis of polar perturbations in the EMP spacetime is technically challenging and is left for future work. Here we adopt a controlled approximation in which linear electromagnetic perturbations are no longer treated as independent dynamical degrees of freedom in the polar sector, but are assumed to arise predominantly from the implicit dependence of the background field on the geometric parameter $q$. 

By linearizing the equation of motion Eq.~(\ref{eom_s}), we obtain the master equation of phantom scalar field perturbations,
\begin{align}
    \square\delta\phi&=-\frac{1}{4}\left(\frac{\partial^2 Z^{-1}}{{\partial\phi}^{2}}F^2\delta\phi+\frac{\partial Z^{-1}}{\partial\phi}\delta F^2\right).\label{linearlized s}
\end{align}
The right-hand side of the above equation consists of the linear term of the phantom scalar field $\delta\phi$ and that of the electromagnetic invariant $\delta F^2$, revealing a coupling between the perturbed phantom scalar field and the perturbed Maxwell field. For the EMP background, the first electromagnetic invariant reads
\begin{align}
    F^2=F^{\mu\nu}F_{\mu\nu}=-2F_{rt}^2=-2\left(\frac{QZ}{r^2+q^2}\right)^2.
\end{align}
Analogous to the metric components in Eq.~(\ref{coefficients_emp}), the linear perturbation of the Maxwell field in the EMP spacetime can be ascribed to an increment of $F_{rt}$ together with the non-vanishing of the other components of the electromagnetic tensor. Importantly, the variation of the electromagnetic invariant is governed solely by $\delta F_{rt}$ at the linear order,
\begin{align}
    \delta F^2=-4F_{rt}\delta F_{rt}=-4F_{rt}\frac{\partial F_{rt}}{\partial q}\frac{\partial q}{\partial\phi}\delta\phi,\label{delta_F^2}
\end{align}
where the contributions from the other components enter only at higher orders. In the final equality we adopt the approximation that the variation $\delta F_{rt}$ is dominated by the implicit dependence of $F_{rt}$ on the geometric parameter $q$, and we hold the total electric charge fixed ($\delta Q=0$), consistent with source-free perturbations that preserve the asymptotic condition. Under this assumption only the $q$-dependence is retained when converting $\delta F_{rt}$ into $\delta\phi$ via the chain rule, which provides an effective closure of the phantom scalar-electromagnetic coupling at the linear order. This closure is an approximation and should be relaxed when polar Maxwell degrees of freedom or charged sources are dynamically important.

The time independence and the spherical symmetry of the EMP metric given by Eqs.~(\ref{EMP_ansatz}) and (\ref{EMP_metric_coef}) imply the decomposition of variables,
\begin{align}
    \delta\phi(t,r,\theta)=\sum_{l,m}\ee^{-\ii\omega t}\frac{\Psi(r)}{\varrho}Y_{lm}(\theta).\label{separation}
\end{align}
Substituting the EMP metric into Eq.~(\ref{linearlized s}) then yields the radial master equation,
\begin{align}
    h^2\frac{\dif^2\Psi}{\dif r^2}+hh'\frac{\dif\Psi}{\dif r}+\left(\omega^2-V\right)\Psi=0,\label{pertur_s_r}
\end{align}
with the effective potential
\begin{align}
    V=h\left(\frac{rh'}{\varrho^{2}}+\frac{q^{2}h}{\varrho^{4}}+\frac{l(l+1)}{\varrho^{2}}-\frac{Q^{2}}{8q^{2}\varrho^2}\left(2rZ'-\varrho^{2}Z''\right)\right),\label{Veff}
\end{align}where a prime denotes the derivative with respect to the radial coordinate. If the last term is omitted in Eq.~(\ref{Veff}), we reduce this effective potential to the case of massless scalar field perturbations in the EMP spacetime as already presented in Ref.~\cite{Ou:2021efv}.


\section{Instabilities of wormholes}
\label{sec:instability}

\subsection{Singularity of the gravito-electromagnetic effective potential}
As we have mentioned before, the coupling function $Z$ is not positive definite and vanishes at
\begin{align}
    \zeta_\pm=&-\frac{q}{\gamma_1}\left(\gamma_2\pm\sqrt{\gamma_1^2+\gamma_2^2}\right),
\end{align}
introducing the divergent behavior in $V_{11}$. In the EMP spacetime, $\zeta_\pm$ mark critical points at which the electromagnetic field transitions from a conventional regime on one side to a phantom-like regime on the other side and can cause the effective potential $V_{11}$ to diverge due to the presence of $Z$ on the denominator. Such phantom-like behavior of the electromagnetic field is not intrinsic to the Maxwell 2-form $F_{\mu\nu}$ itself, but is induced entirely by the nonminimal coupling $Z^{-1}$ between the Maxwell field and the phantom scalar field in the Lagrangian. The primary source of the singular behavior of $V_{11}$ is the phantom scalar field: The singular behavior of the gravito-electromagnetic effective potential is a direct consequence of their coupling to the phantom scalar field. If the phantom scalar field were absent, the EMP action would reduce to the Einstein--Maxwell action, and the spacetime would revert to the stable RN solution. In this limit, the singular behavior of the gravito-electromagnetic effective potential would cease to exist.

In the Type I black hole scenario, these divergences are strictly confined within the Cauchy horizon ($r_->\zeta_+$), a condition equivalent to
\begin{align}
    8\gamma_2q^2  +\gamma_1\gamma_2Q^2 >\gamma_1 \sqrt{\gamma_2^2Q^4-16\gamma_1q^2Q^2-64q^4}+8 q^2\sqrt{\gamma_1^2 + \gamma_2^2}.\label{instability1}
\end{align}
Through two consecutive squaring operations applied to both sides, the inequality Eq.~(\ref{instability1}) transforms into the manifestly satisfied condition,
\begin{align}
    \left(\gamma_1^2+\gamma_2^2\right)\left(8q^2+\gamma_1 Q^2\right)^2>0.
\end{align}
Hence, the divergence points of $V_{11}$ are proven to be hidden behind the Cauchy horizon, which indicates the stability of EMP black holes\footnote{Here we address that the linear stability of the exterior region of Type I black holes is studied in analogy with that of RN spacetimes; we note that the interior region of RN black holes, including the extremal case as well, has been shown to be unstable under linear perturbations, see Ref.~\cite{Dotti:2010} and the references therein.}. Nevertheless, as the electric charge increases beyond the extremal value, the black hole undergoes a transition to a wormhole configuration, where singularities of $V_{11}$ become exposed to the outer spacetime, violating cosmic censorship and rendering the configuration physically inadmissible.

\subsection{Finite difference method}
The finite difference method is widely used to simulate the time evolution of perturbations~\cite{Ou:2021efv,Yang:2024prm}. For this purpose, we grid the spacetime coordinates $t=i\Delta t$ and $r_*=j\Delta r_*$, where $i$ and $j$ are both integers. Taking the step length $\Delta t$ and $\Delta r_*$ to be constant, $\mathbf{H}\left(t,r_*\right)$ and $\mathbf{V}\left(t,r_*\right)$ can be abbreviated as $\mathbf{H}\left(i,j\right)$ and $\mathbf{V}\left(i,j\right)$. The master equation Eq.~(\ref{master_eq}) can therefore be expressed as 
\begin{eqnarray}
& &-\frac{\mathbf{H}\left(i+1,j\right)-2\mathbf{H}\left(i,j\right)+\mathbf{H}\left(i-1,j\right)}{\Delta t^2}
+\frac{\mathbf{H}\left(i,j+1\right)-2\mathbf{H}\left(i,j\right)+\mathbf{H}\left(i,j-1\right)}{\Delta r_*^2}\nonumber \\
& &-\mathbf{V}\left(j\right)\mathbf{H}\left(i,j\right)=0.\label{finite difference}
\end{eqnarray}
We consider the initial Gaussian distribution,
\begin{align}
    \mathbf{H}\left(t=0,r_*\right)=\left(\begin{array}{c}\ee^{-\frac{\left(r_*-a\right)^2}{2b^2}}\\0\end{array}\right),\qquad\mathbf{H}\left(t<0,r_*\right)=\mathbf{0},
\end{align}
where $a$ and $b$ are the median and width of the initial wave packet, respectively, and then calculate the time evolution of the perturbation via the iteration scheme,
\begin{align}
    \mathbf{H}\left(i+1,j\right)=-\mathbf{H}\left(i-1,j\right)+\left[2-2\frac{\Delta t^2}{\Delta r_*^2}-\Delta t^2 \mathbf{V}\left(j\right)\right]\mathbf{H}\left(i,j\right)+\frac{\Delta t^2}{\Delta r_*^2}\left[\mathbf{H}\left(i,j+1\right)+\mathbf{H}\left(i,j-1\right)\right].
\end{align}
For the phantom scalar field perturbation, we can simply replace $\mathbf{H}$ and $\mathbf{V}$ in Eq. (\ref{finite difference}) by $\Psi$ and $V$ in Eq. (\ref{pertur_s_r}). The initial condition is
\begin{align}
    \Psi(t=0,r_*)=\ee^{-\frac{\left(r_*-a\right)^2}{2b^2}},\qquad \Psi(t<0,r_*)=0.
\end{align}To satisfy the von Neumann stability condition, $\Delta t/\Delta r_*<1$, we set $\Delta t=0.1$, $\Delta r_*=0.2$, $a=20$, and $b=1$.

\subsection{Divergence of phantom scalar field perturbations}
A previous study reported~\cite{Ou:2021efv} that EMP wormholes remain stable under massless scalar field perturbations. However, the phantom scalar field, serving as the matter source in the EMP spacetime, carries negative kinetic energy and can therefore induce spacetime instabilities. 

\begin{figure}[!ht]
	\centering
	\begin{subfigure}[b]{0.49\textwidth}
		\centering
		\includegraphics[width=\textwidth]{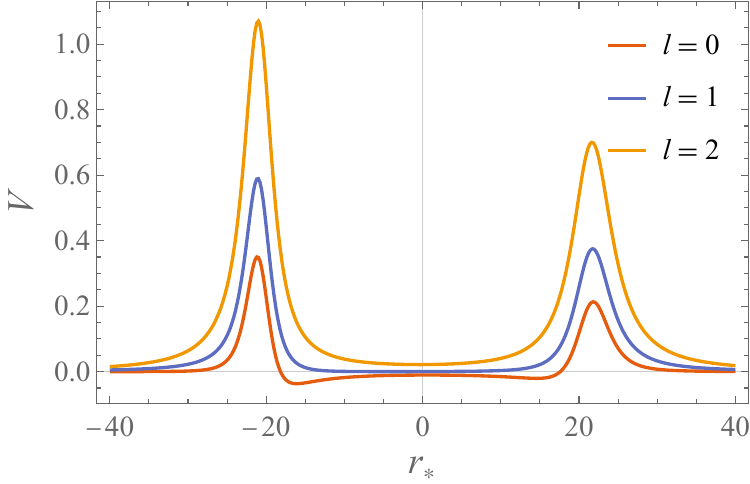}
		\caption{~Effective potential.}
	\end{subfigure}
	\begin{subfigure}[b]{0.49\textwidth}
		\centering
		\includegraphics[width=\textwidth]{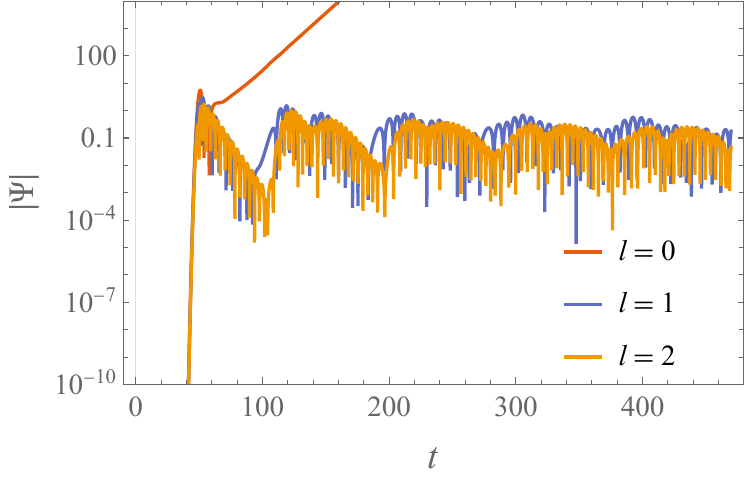}
		\caption{~Time profile.}
	\end{subfigure}
	\captionsetup{width=\textwidth}
	\caption{The effect of the azimuthal number $l$ of EMP wormholes on the effective potential and the time profile of the phantom scalar field perturbation  with $q=1.6$, $Q=1$, $\gamma_1=-10$ and $\gamma_2=2$.}
	\label{fig:wormhole}
\end{figure}

Fig.~\ref{fig:wormhole} shows how the azimuthal number $l$ affects the effective potential and time domain waveform for phantom scalar field perturbations. For $l=0$, the effective potential contains a negative region, causing the waveform to grow exponentially after the initial ringing. As $l$ increases, this region disappears and the effective potential develops a double-peak structure, yielding echo signals in the waveform. These echoes follow the patterns remarkably similar to those of a minimally coupled massless scalar field in Ref.~\cite{Ou:2021efv}, and will not be further discussed here. Instead, we focus on the divergence of phantom scalar field perturbations and the singular behavior of $V_{11}$ as mentioned above, both of which indicate the instability of the EMP wormhole. Here we proceed with the following possible explanations:
\begin{itemize}
	\item As the electric charge increases, Type I black holes go to extremal and turn into wormholes. However, extremal black holes are unattainable in general relativity, implying that neither extremal Type I black holes nor wormholes in the EMP theory are likely to be physically realized.
	\item The divergence of $V_{11}$ can be resolved if the electromagnetic field perturbation $H_1$ vanishes at $r=\zeta_{\pm}$. 
	\item EMP wormholes are unstable under gravitational field perturbations because of the coupling between the phantom scalar field and the Maxwell field.
	\item The divergence of $V_{11}$ may arise from limitations of linearization procedure or from an inadequate choice of perturbation variables, rather than an essential singularity.
\end{itemize}

The above discussion is based on the linear perturbation theory and does not determine the nonlinear outcome. However, the studies of  wormholes supported by phantom scalar fields indicate that the linear instability typically persists nonlinearly, with the spacetime either expanding or collapsing to a black hole depending on the initial perturbation~\cite{Gonzalez:2008wd,Gonzalez:2008xk,Gonzalez:2009hn}. Whether EMP wormholes follow the same fate or not needs deeper investigations in future.

As a summary to this section, we have revealed two complementary signs of EMP wormhole instability: The exponential growth of $l=0$ phantom scalar field perturbations and a divergence in the gravito-electromagnetic effective potential --- while the former can be resolved by suitably tuning spacetime parameters, the latter persists. It is the presence of the phantom scalar field and its coupling with the Maxwell field responsible for the instability of EMP wormholes. During the wormhole--black hole transition, the event horizon forms in such a way that it hides these singularities behind the Cauchy horizon. Consequently, although the theory admits wormhole solutions, only black hole configurations with a complete horizon structure are physically viable and stable. A more detailed investigation of the underlying mechanisms and potential remedies remains an open question for future work.
\section{Numerical results}
\label{sec:numerical}
\subsection{Quasinormal modes of Type I black holes}
In the previous section, the instability of wormholes as demonstrated indirectly supports the stability of black holes, where the latter will be explored in great detail in this section. We will solve the master equation Eq.~(\ref{master_eq}) numerically to acquire QNFs of Type I black holes. The boundary conditions for the QNM problem are pure ingoing waves at the event horizon, 
\begin{align}
    \Psi,\ \mathbf{H}\sim\ee^{-\ii\omega r_*},\qquad r_*\to-\infty,
\end{align}
and pure outgoing waves at the spatial infinity,
\begin{align}
    \Psi,\ \mathbf{H}\sim\ee^{\ii\omega r_*},\qquad r_*\to\infty,
\end{align}
where the tortoise coordinate for Type I black holes is given by
\begin{align}
    r_{*}=r+\frac{1}{2\beta}\ln{\frac{\abs{r-r_{+}}}{2M}}-\frac{\alpha}{2\beta}\ln{\frac{\abs{r-r_{-}}}{2M}},
\end{align}
with
\begin{align}
    \beta=\frac{r_{+}-r_{-}}{2\left(r_{+}^{2}+q^{2}\right)},\qquad\alpha=\frac{r_{-}^2+q^2}{r_{+}^{2}+q^2}.
\end{align}
The boundary conditions can be approximated by plane waves because the EMP spacetime is asymptotically flat and the perturbation equations take a Schr{\"o}dinger-like form. The three perturbing fields exhibit similar but not identical asymptotic behavior at the boundaries. To make the distinct asymptotics of each field explicit, we present their Frobenius expansions,
\begin{align}
    \psi_\mathrm{ps}&=\ee^{\varrho r}(r-r_{+})^{\tfrac{\varrho}{2\beta} }(r-r_{-})^{-\tfrac{\alpha\varrho}{2\beta} }{\ee}^{-2\varrho(r-r_+)}(r-r_{+}+1)^{-\tfrac{\varrho}{\beta}(1-\alpha)}\sum_n a^\mathrm{ps}_n\Psi^n,\\
    \psi_\mathrm{em}&=\ee^{\varrho r}(r-r_{+})^{\tfrac{\varrho}{2\beta} }(r-r_{-})^{-\tfrac{\alpha\varrho}{2\beta} }{\ee}^{-2\varrho(r-r_+)}(r-r_{+}+1)^{-\tfrac{\varrho}{\beta}(1-\alpha)}\sum_n a^\mathrm{em}_n H_1^n,\\
    \psi_\mathrm{g}&=r^{-1}\ee^{\varrho r}(r-r_{+})^{\tfrac{\varrho}{2\beta} }(r-r_{-})^{-\tfrac{\alpha\varrho}{2\beta} }{\ee}^{-2\varrho(r-r_+)}(r-r_{+}+1)^{-\tfrac{\varrho}{\beta}(1-\alpha)}\varrho Z^{-1/2}\sum_n a^\mathrm{g}_n H_2^n,
\end{align}
where $a_{n}$ denotes the amplitude, and $\psi_\mathrm{ps}$, $\psi_\mathrm{em}$,  and $\psi_\mathrm{g}$ correspond to the phantom scalar field, the electromagnetic field and the gravitational field, respectively.

Compared with the Type II solution, we are more interested in the Type I scenario, where the latter recover RN black holes of general relativity if the condition in Eq.~(\ref{limit_to_GR}) is adopted. We first examine the scalar charge dependence of the QNMs for Type I black holes in the EMP spacetime and impose $\gamma_1=1/4$, leaving $q$, $Q$, and $\gamma_2$ as three free parameters. We next consider the expression of the mass and electric charge, see Eq. (\ref{mass}), and set $\gamma_2=2q$ for convenience,\footnote{$\gamma_2$ must be proportional to $q$ in order to recover the RN solution. Our choice of $\gamma_2=2q$ might be of arbitrariness to a certain extent, but such an arbitrariness can be eliminated once the QNFs are rendered dimensionless by the multiplication with the mass $M$.} so that the specific electric charge is the inverse of $Q$, i.e.,
\begin{align}
    \frac{Q_\mathrm{e}}{M}=\frac{1}{Q}.
\end{align}
After assigning the specific electric charge, we can then vary $q$ over the interval $(0,q_\mathrm{ext}]$. The lower bound $q=0$ follows from the quantum-gravity consistency; the upper bound $q_\mathrm{ext}$ denotes the extremal value of black hole configurations, which is fixed in terms of the stability of black hole configurations and the instability of wormhole configurations,
\begin{align}
    q_\mathrm{ext}=\frac{Q}{4} \sqrt{Q^2-1}=\frac{M}{4 Q_\mathrm{e}}\sqrt{\left(\frac{M}{Q_\mathrm{e}}\right)^2-1}.
\end{align}

The QNFs are determined through two complementary methodologies: The extraction from the finite difference time domain profiles via the Prony method and the computation via a matrix-valued direct integration approach; see Refs.~\cite{Berti:2007,Konoplya:2011qq} for the Prony method and Ref.~\cite{Pani:2013pma} for the direct integration method. We represent the fundamental QNFs of the phantom scalar field, the electromagnetic field and the gravitational field perturbations of Type I black holes with $Q_\mathrm{e}/M=2/5$ and $q_\mathrm{ext}=\frac{5\sqrt{21}}{16}$ in Table~\ref{tab:q}. The azimuthal number is set to be $l=2$, corresponding to the lowest nontrivial gravitational quadrupole mode.

\begin{center}
	\begin{table}[!ht]
		\resizebox{\textwidth}{!}{
			\begin{tabular}{|c|c|c|c|c|c|c|c|c|c|}
				\hline
				$q$&\multicolumn{2}{c|}{Phantom scalar field perturbation}&\multicolumn{2}{c|}{Electromagnetic field perturbation}&\multicolumn{2}{c|}{Gravitational field perturbation}   \\ \hline
				& Direct integration                  &  Prony        & Direct integration                  &  Prony            & Direct integration                &  Prony   \\ \hline
				&$\omega_{\mathrm{R}}M$~~~~~~~$\omega_{\mathrm{I}}M$&$\omega_{\mathrm{R}}M$~~~~~~~$\omega_{\mathrm{I}}M$&$\omega_{\mathrm{R}}M$~~~~~~~$\omega_{\mathrm{I}}M$&$\omega_{\mathrm{R}}M$~~~~~~~$\omega_{\mathrm{I}}M$&$\omega_{\mathrm{R}}M$~~~~~~~$\omega_{\mathrm{I}}M$&$\omega_{\mathrm{R}}M$~~~~~~~$\omega_{\mathrm{I}}M$\\ \hline
				0.001   &0.59187~~~-0.10237&0.59191~~~-0.10239&0.47993~~-0.096442&0.47994~~-0.096469&0.37844~~-0.089398&0.37847~~-0.089422\\ \hline
				0.2 &0.59194~~~-0.10215&0.59198~~~-0.10217&0.48187~~-0.096421&0.48192~~-0.096457&0.37810~~-0.089208&0.37813~~-0.089232\\ \hline
				0.4 &0.59213~~~-0.10146&0.59217~~~-0.10149&0.48821~~-0.096495&0.48828~~-0.096525&0.37725~~-0.088654&0.37728~~-0.088687\\ \hline
				0.6 &0.59245~~~-0.10028&0.59249~~~-0.10031&0.49874~~-0.096263&0.49879~~-0.096305&0.37614~~-0.087714&0.37617~~-0.087736\\ \hline
				0.8 &0.59288~~-0.098547&0.59293~~-0.098569&0.51290~~-0.095261&0.51297~~-0.095300&0.37497~~-0.086303&0.37500~~-0.086323\\ \hline
				1   &0.59339~~-0.096145&0.59343~~-0.096167&0.53111~~-0.093153&0.53113~~-0.093161&0.37391~~-0.084267&0.37392~~-0.084265\\ \hline
				1.2 &0.59387~~-0.092929&0.59392~~-0.092950&0.55485~~-0.089156&0.55493~~-0.089197&0.37302~~-0.081359&0.37304~~-0.081379\\ \hline
				1.4 &0.59412~~-0.088730&0.59416~~-0.088751&0.58734~~-0.080641&0.58743~~-0.080682&0.37208~~-0.077231&0.37210~~-0.077250\\ \hline
				1.43&0.59412~~-0.088029&0.59416~~-0.088029&0.59332~~-0.078506&0.59339~~-0.078546&0.37190~~-0.076496&0.37192~~-0.076515\\ \hline	
		\end{tabular}}\\
		\caption{Fundamental QNMs of the phantom scalar field perturbation, the electromagnetic field perturbation and the gravitational field  perturbation for Type I black holes with parameters $\gamma_1=1/4$, $\gamma_2=2q$ and $Q_\mathrm{e}/M=2/5$, computed via the direct integration method and the Prony method for a varying scalar charge $q$. The azimuthal number $l$ is fixed at $l=2$.}
		\label{tab:q}
	\end{table}
\end{center}

A critical consistency check emerges in the vanishing scalar charge limit, where both the master equations Eqs.~(\ref{electromagnetic master euqation}) and (\ref{gravito master euqation}) as well as the corresponding QNFs reproduce the case of axial (odd-parity) gravito-electromagnetic field perturbations of RN black holes. This concordance (notably, the cases of three field perturbations at $q=0.001$ in Table~\ref{tab:q} match the RN predictions) provides a robust validation of our calculation. Moreover, the agreement between the direct integration and Prony results confirms the validity of the numerical methods. Although the direct integration method is highly efficient for simple perturbation equations like the phantom scalar field perturbation equation, see Eq.~(\ref{pertur_s_r}), it becomes time-consuming for  complex ones like the gravito-electromagnetic coupled perturbation equations, see Eqs.~(\ref{electromagnetic master euqation}) and (\ref{gravito master euqation}). Accordingly, we employ the direct integration method for phantom scalar field perturbations and use the finite difference method combined with the Prony method for gravito-electromagnetic field perturbations.

\begin{figure}[!ht]
	\centering
	\begin{subfigure}[b]{0.49\textwidth}
		\centering
		\includegraphics[width=\textwidth]{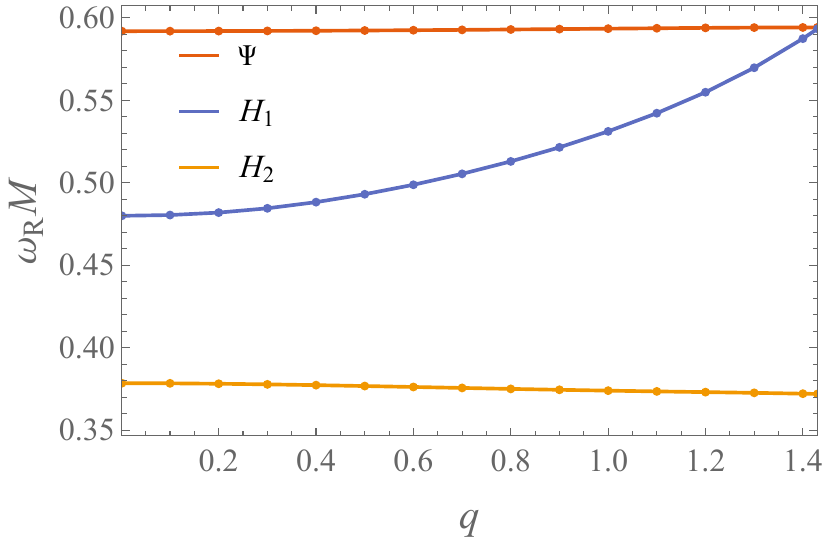}
		\caption{~Real parts of QNFs.}
	\end{subfigure}
	\begin{subfigure}[b]{0.49\textwidth}
		\centering
		\includegraphics[width=\textwidth]{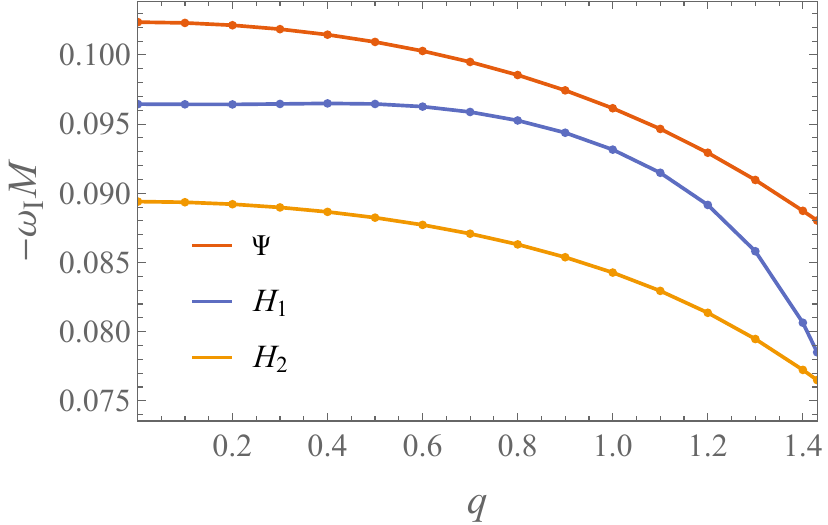}
		\caption{~Imaginary parts of QNFs.}
	\end{subfigure}
	\captionsetup{width=\textwidth}
	\caption{The effect of the scalar charge $q$ of Type I black holes on the QNFs with $\gamma_1=1/4$, $\gamma_2=2q$, $Q_\mathrm{e}/M=2/5$ and $l=2$.}
	\label{fig:QNF_q}
\end{figure}

Fig.~\ref{fig:QNF_q} demonstrates that the scalar charge $q$ induces distinct spectral shifts in QNFs for different field perturbations.
\begin{itemize}
\item
For the phantom scalar and electromagnetic field perturbations, the real frequency $\omega_\mathrm{R}$ increases when $q$ grows, whereas it decreases slightly for the gravitational field perturbation. 
\item
The imaginary part $\omega_\mathrm{I}$ of QNFs grows monotonically for both the phantom scalar and gravitational field perturbations as $q$ rises, suggesting that a larger scalar charge leads to a longer lived ringdown signal. By contrast, it exhibits a non-monotonic dependence for the electromagnetic field perturbation: $\omega_\mathrm{I}$ initially decreases at a small scalar charge $q$ and then increases at a large $q$.
\end{itemize}

Next we investigate how the relative contributions of the phantom scalar and electromagnetic field perturbations affect the QNMs. By rescaling $q$ and $Q_\mathrm{e}$ by the mass $M$, i.e.,  $\tilde{q}=q/M$ and $\tilde{Q}_\mathrm{e}=Q_\mathrm{e}/M$, we further define the generalized specific charge,
\begin{align}
    \eta=\frac{\sqrt{q^2+Q_\mathrm{e}^2}}{M}=\sqrt{\tilde{q}^2+\tilde{Q}_\mathrm{e}^2},
\end{align}
where $\eta=1$ corresponds to extremal black holes. We introduce a mixing angle between the scalar and electric charges by
\begin{align}
    \tan{\vartheta}\equiv\frac{{q}}{{Q}_\mathrm{e}}=\frac{\tilde{q}}{\tilde{Q}_\mathrm{e}},
\end{align}
where $\vartheta\in[0,\pi/2]$, and express the dimensionless charges using $\eta$ and $\vartheta$,
\begin{align}
    \tilde{q}=\eta\sin{\vartheta},\qquad\tilde{Q}_\mathrm{e}=\eta\cos{\vartheta}.
\end{align}
If $\eta$ is fixed but $\vartheta$ varied, we can keep the horizon radius unchanged but adjust the 
relative contributions of the scalar and electric charges.

\begin{figure}[!ht]
	\centering
	\begin{subfigure}[b]{0.49\textwidth}
		\centering
		\includegraphics[width=\textwidth]{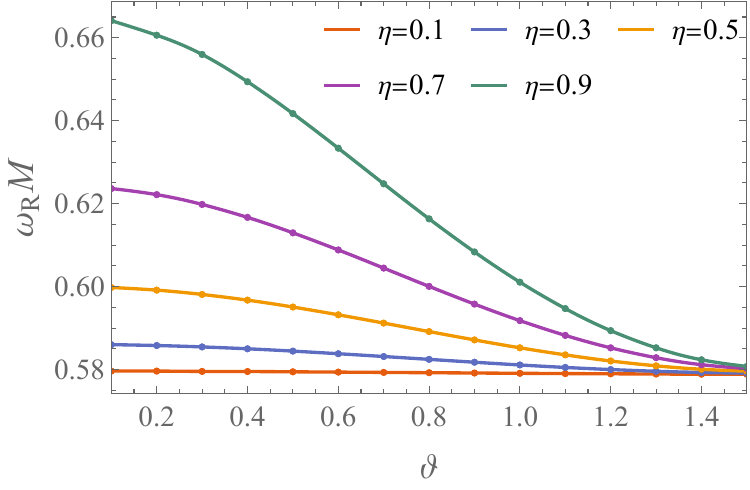}
		\caption{~Real parts of QNFs.}
	\end{subfigure}
	\begin{subfigure}[b]{0.49\textwidth}
		\centering
		\includegraphics[width=\textwidth]{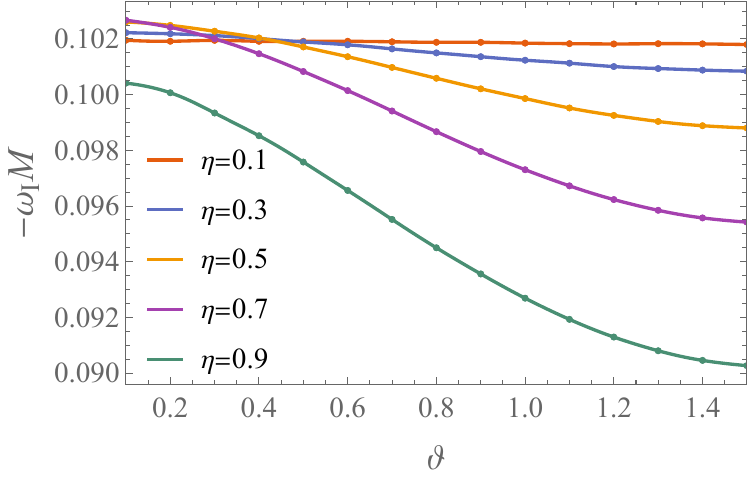}
		\caption{~Imaginary parts of QNFs.}
	\end{subfigure}
	\captionsetup{width=\textwidth}
	\caption{The effect of the mixing angle $\vartheta$ of Type I black holes on the fundamental QNFs of the phantom scalar field perturbation with $l=2$.}
	\label{fig:QNF_PS}
\end{figure}

\begin{figure}[!ht]
	\centering
	\begin{subfigure}[b]{0.49\textwidth}
		\centering
		\includegraphics[width=\textwidth]{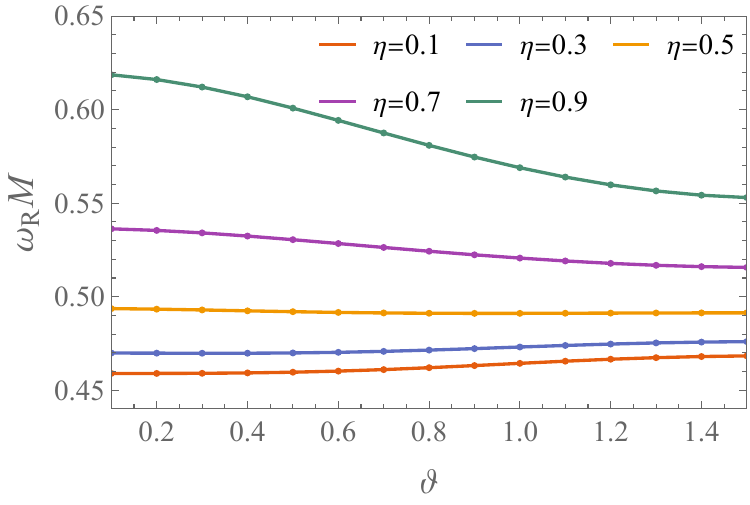}
		\caption{~Real parts of QNFs.}
	\end{subfigure}
	\begin{subfigure}[b]{0.49\textwidth}
		\centering
		\includegraphics[width=\textwidth]{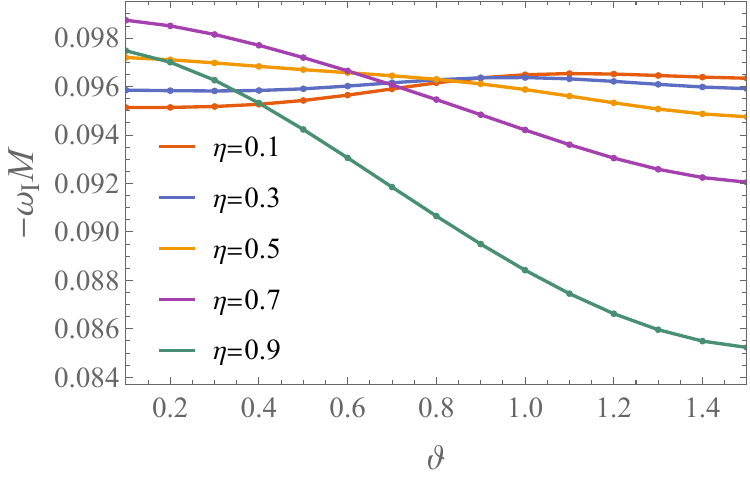}
		\caption{~Imaginary parts of QNFs.}
	\end{subfigure}
	\captionsetup{width=\textwidth}
	\caption{The effect of the mixing angle $\vartheta$ of Type I black holes on the fundamental QNFs of the electromagnetic field perturbation with $l=2$.}
	\label{fig:QNF_EM}
\end{figure}

\begin{figure}[!ht]
	\centering
	\begin{subfigure}[b]{0.49\textwidth}
		\centering
		\includegraphics[width=\textwidth]{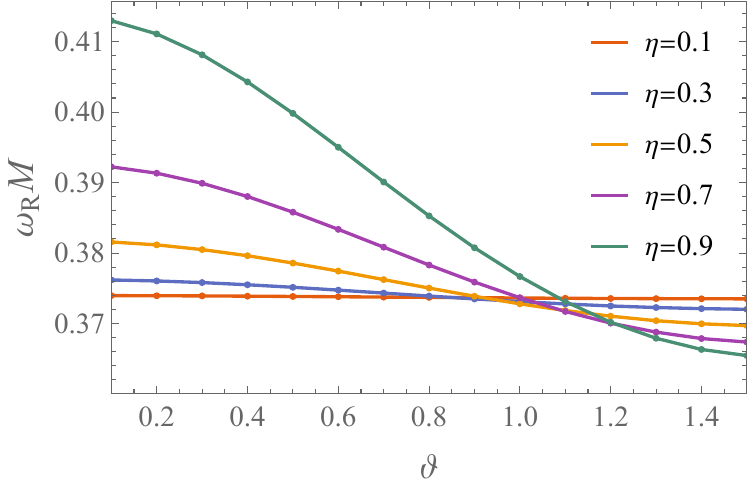}
		\caption{~Real parts of QNFs.}
	\end{subfigure}
	\begin{subfigure}[b]{0.49\textwidth}
		\centering
		\includegraphics[width=\textwidth]{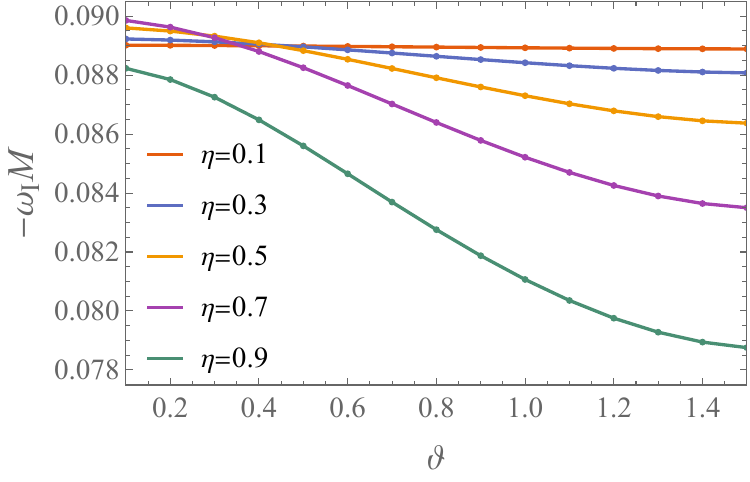}
		\caption{~Imaginary parts of QNFs.}
	\end{subfigure}
	\captionsetup{width=\textwidth}
	\caption{The effect of the mixing angle $\vartheta$ of Type I black holes on the fundamental QNFs of the gravitational field perturbation with $l=2$.}
	\label{fig:QNF_Grav}
\end{figure}

Figs.~\ref{fig:QNF_PS}-\ref{fig:QNF_Grav} illustrate how the fundamental QNFs depend on the charge composition for the phantom scalar field, electromagnetic field, and gravitational field perturbations of Type I black holes, respectively. The real frequency and damping rate of the perturbed phantom scalar and gravitational fields, see Figs.~\ref{fig:QNF_PS} and \ref{fig:QNF_Grav}, decrease monotonically as the mixing angle $\vartheta$ increases, in particular, the decreasing amplitudes increase for larger values of the generalized specific charge $\eta$. For the perturbed electromagnetic field, see Fig.~\ref{fig:QNF_EM}, the real frequency and damping rate show a different tendency:  $\omega_{\mathrm{R}}$ increases monotonically as the mixing angle $\vartheta$ increases at a low $\eta$ but decreases monotonically at a high $\eta$; the damping rate rises when the mixing angle is small at a low $\eta$, but it decreases monotonically when the mixing angle increases at a high $\eta$. The values of QNFs are given in Tables \ref{tab:QNF_PS}-\ref{tab:QNF_Grav}. As the parameters approach the transition values from the black-hole side, our numerical survey shows that the quasinormal frequencies vary smoothly and continuously near the transition, suggesting that the spacetime geometry changes rather gently.

\begin{center}
	\begin{table}[!ht]
		\resizebox{\textwidth}{!}{
			\begin{tabular}{|c|c|c|c|c|c|}
				\hline
				\multirow{2}{*}{\diagbox{$\vartheta$}{$\eta$}}&$0.1$&$0.3$&$0.5$&$0.7$&$0.9$\\
				\cline{2-6}
				&   $\omega_\mathrm{R}M$~~~~~~$\omega_\mathrm{I}M$&$\omega_\mathrm{R}M$~~~~~~$\omega_\mathrm{I}M$&$\omega_\mathrm{R}M$~~~~~~$\omega_\mathrm{I} M$&$\omega_\mathrm{R} M$~~~~~~~$\omega_\mathrm{I}M$&$\omega_\mathrm{R}M$~~~~~~$\omega_\mathrm{I}M$\\
				\hline
				0.1&0.57963~~~-0.10196&0.58600~~~-0.10223&0.59979~~~-0.10261&0.62364~~~-0.10267&0.66411~~~-0.10041\\
				\hline
				0.3&0.57954~~~-0.10195&0.58545~~~-0.10212&0.59810~~~-0.10228&0.61981~~~-0.10201&0.65597~~-0.099341\\
				\hline
				0.5&0.57945~~~-0.10192&0.58444~~~-0.10189&0.59507~~~-0.10171&0.61296~~~-0.10083&0.64171~~-0.097580\\
				\hline
				0.7&0.57931~~~-0.10190&0.58314~~~-0.10164&0.59120~~~-0.10098&0.60446~~-0.099413&0.62478~~-0.095520\\
				\hline
				0.9&0.57916~~~-0.10188&0.58176~~~-0.10136&0.58714~~~-0.10021&0.60005~~-0.098671&0.61633~~-0.094501\\
				\hline
				1.1&0.57903~~~-0.10183&0.58050~~~-0.10113&0.58354~~-0.099522&0.59179~~-0.097304&0.60107~~-0.092691\\
				\hline
				1.3&0.57892~~~-0.10183&0.57958~~~-0.10094&0.58089~~-0.099038&0.58285~~-0.095847&0.58524~~-0.090803\\
				\hline
				1.5&0.57888~~~-0.10180&0.57912~~~-0.10084&0.57960~~-0.098802&0.58024~~-0.095422&0.58076~~-0.090269\\
				\hline
				
		\end{tabular}}\\
		\caption{Fundamental QNMs of the phantom scalar field perturbation for Type I black holes with $\gamma_1=1/4$, where the direct integration method is used for different values of $\vartheta$ and $\eta$. The azimuthal number $l$ is set to be $l=2$.}
		\label{tab:QNF_PS}
	\end{table}
\end{center}
\begin{center}
	\begin{table}[!ht]
		\resizebox{\textwidth}{!}{
			\begin{tabular}{|c|c|c|c|c|c|}
				\hline
				\multirow{2}{*}{\diagbox{$\vartheta$}{$\eta$}}&$0.1$&$0.3$&$0.5$&$0.7$&$0.9$\\
				\cline{2-6}
				&   $\omega_\mathrm{R}M$~~~~~~$\omega_\mathrm{I}M$&$\omega_\mathrm{R}M$~~~~~~$\omega_\mathrm{I}M$&$\omega_\mathrm{R}M$~~~~~~$\omega_\mathrm{I} M$&$\omega_\mathrm{R} M$~~~~~~~$\omega_\mathrm{I}M$&$\omega_\mathrm{R}M$~~~~~~$\omega_\mathrm{I}M$\\
				\hline
				0.1&0.45898~~~-0.095134&0.46989~~~-0.095854&0.49364~~~-0.097200&0.53630~~~-0.098740&0.61865~~~-0.097479\\
				\hline
				0.3&0.45906~~~-0.095178&0.46978~~~-0.095819&0.49293~~~-0.096978&0.53413~~~-0.098152&0.61209~~~-0.096264\\
				\hline
				0.5&0.45964~~~-0.095427&0.46994~~~-0.095908&0.49199~~~-0.096700&0.53049~~~-0.097197&0.60083~~~-0.094229\\
				\hline
				0.7&0.46103~~~-0.095904&0.47078~~~-0.096157&0.49132~~~-0.096449&0.52634~~~-0.096069&0.58754~~~-0.091852\\
				\hline
				0.9&0.46319~~~-0.096360&0.47225~~~-0.096361&0.49110~~~-0.096113&0.52240~~~-0.094838&0.57467~~~-0.089498\\
				\hline
				1.1&0.46553~~~-0.096538&0.47394~~~-0.096320&0.49117~~~-0.095609&0.51912~~~-0.093601&0.56397~~~-0.087451\\
				\hline
				1.3&0.46741~~~-0.096458&0.47530~~~-0.096093&0.49130~~~-0.095073&0.51680~~~-0.092586&0.55655~~~-0.085962\\
				\hline
				1.5&0.46836~~~-0.096345&0.47599~~~-0.095914&0.49137~~~-0.094753&0.51567~~~-0.092051&0.55301~~~-0.085230\\
				\hline
				
		\end{tabular}}\\
		\caption{Fundamental QNMs of the electromagnetic field perturbation for Type I black holes with $\gamma_1=1/4$, where the Prony method is used for different values of $\vartheta$ and $\eta$. The azimuthal number $l$ is set to be $l=2$.}
		\label{tab:QNF_EM}
	\end{table}
\end{center}

\begin{center}
	\begin{table}[!ht]
		\resizebox{\textwidth}{!}{
			\begin{tabular}{|c|c|c|c|c|c|}
				\hline
				\multirow{2}{*}{\diagbox{$\vartheta$}{$\eta$}}&$0.1$&$0.3$&$0.5$&$0.7$&$0.9$\\
				\cline{2-6}
				&   $\omega_\mathrm{R}M$~~~~~~$\omega_\mathrm{I}M$&$\omega_\mathrm{R}M$~~~~~~$\omega_\mathrm{I}M$&$\omega_\mathrm{R}M$~~~~~~$\omega_\mathrm{I} M$&$\omega_\mathrm{R} M$~~~~~~~$\omega_\mathrm{I}M$&$\omega_\mathrm{R}M$~~~~~~$\omega_\mathrm{I}M$\\
				\hline
				0.1&0.37396~~~-0.089014&0.37618~~~-0.089226&0.38157~~~-0.089602&0.39223~~~-0.089855&0.41297~~~-0.088231\\
				\hline
				0.3&0.37392~~~-0.089004&0.37581~~~-0.089129&0.38049~~~-0.089322&0.38990~~~-0.089269&0.40813~~~-0.087251\\
				\hline
				0.5&0.37384~~~-0.088985&0.37514~~~-0.088959&0.37858~~~-0.088830&0.38580~~~-0.088251&0.39982~~~-0.085599\\
				\hline
				0.7&0.37375~~~-0.088962&0.37432~~~-0.088749&0.37623~~~-0.088225&0.38083~~~-0.087021&0.39008~~~-0.083693\\
				\hline
				0.9&0.37366~~~-0.088938&0.37349~~~-0.088527&0.37387~~~-0.087598&0.37588~~~-0.085784&0.38075~~~-0.081874\\
				\hline
				1.1&0.37358~~~-0.088915&0.37278~~~-0.088320&0.37184~~~-0.087027&0.37171~~~-0.084699&0.37312~~~-0.080355\\
				\hline
				1.3&0.37353~~~-0.088897&0.37227~~~-0.088158&0.37040~~~-0.086591&0.36878~~~-0.083899&0.36791~~~-0.079279\\
				\hline
				1.5&0.37350~~~-0.088887&0.37202~~~-0.088074&0.36971~~~-0.086369&0.36737~~~-0.083501&0.36545~~~-0.078755\\
				\hline
				
		\end{tabular}}\\
		\caption{Fundamental QNMs of the gravitational field perturbation for Type I black holes with $\gamma_1=1/4$, where the Prony method is used for different values of $\vartheta$ and $\eta$. The azimuthal number $l$ is set to be $l=2$.}
		\label{tab:QNF_Grav}
	\end{table}
\end{center}

\subsection{Observability of quasinormal mode deviations}
The spectral differences between Type I black holes and their RN counterparts, as shown in Table~\ref{tab:q}, raise the practical question of whether gravitational wave observations can distinguish the two classes of black holes. A common and heuristic resolvability test for two damped sinusoids is given by the Rayleigh criterion~\cite{Berti:2005ys},
\begin{align}
    (f_1-f_2)\tau>1,
\end{align}
where the oscillation frequency $f$ and the damping time $\tau$ are related to the complex QNF by
\begin{align}
    f=\frac{\omega_\mathrm{R}}{2\pi},\qquad\tau=\frac{1}{\abs{\omega_\mathrm{I}}}.
\end{align}
The quality factor $\mathscr{Q}$ is defined as 
\begin{align}
    \mathscr{Q}\equiv\pi f\tau =\frac{\omega_{\mathrm{R}}}{2\abs{\omega_{\mathrm{I}}}},
\end{align}
which characterizes the number of oscillation cycles before the signal decays.

To assess resolvability quantitatively, one must compare the QNM shifts to the measurement uncertainties in frequency $\sigma_f$ and damping time $\sigma_\tau$. To the leading order and for a large $\mathscr{Q}$, these error scales take the forms~\cite{Berti:2005ys},
\begin{align}
    \sigma_f\simeq\frac{1}{\sqrt{2}\pi\tau\iota},\qquad\sigma_\tau\simeq\frac{2\tau}{\iota},
\end{align}
where $\iota$ denotes the signal-to-noise ratio (SNR) of the ringdown signal. Accordingly, we define the observability measurements for frequency and damping-time deviations as
\begin{align}
    \mathcal{O}_f\equiv\frac{\abs{\Delta f}}{\sigma_f}=\frac{\abs{\Delta f}}{f}\frac{f}{\sigma_f}=\sqrt{2}\pi\iota\mathscr{Q}\frac{\abs{\Delta f}}{f},\qquad\mathcal{O}_\tau\equiv\frac{\abs{\Delta\tau}}{\sigma_\tau}=\frac{\abs{\Delta\tau}}{\tau}\frac{\tau}{\sigma_\tau}=\frac{\iota}{2}\frac{\abs{\Delta\tau}}{\tau},
\end{align}
where $\Delta f$ and $\Delta\tau$ denote differences between the EMP and RN values. A deviation is considered to be observable when the corresponding ratio exceeds unity, i.e., $\mathcal{O}_f>1$ or $\mathcal{O}_\tau>1$.

Fig.~\ref{fig:observability} summarizes our numerical estimates in the following three points:

\begin{figure}[!ht]
    \centering
    \includegraphics[width=0.75\linewidth]{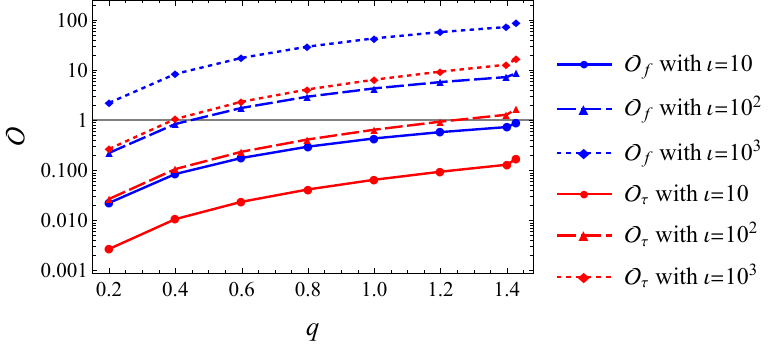}
    \caption{The observability of deviations in the fundamental gravitational QNM spectra ($l=2,n=0$) between Type I black holes and RN black holes. The blue and red lines represent the observability of frequency $\mathcal{O}_f$ and the observability of damping time $\mathcal{O}_\tau$ for different values of the scalar charge $q$, with circular, triangular and diamond markers corresponding to SNRs of $\iota=10$, $10^2$ and $10^3$, respectively. The horizontal line at $\mathcal{O}=1$ indicates the detection threshold.}
    \label{fig:observability}
\end{figure}

\begin{itemize}
    \item For a modest network SNR $\iota=10$, one representative of most current detections~\cite{LIGOScientific:2025slb}, neither the frequency nor the damping-time shift between Type I and RN gravitational QNMs is resolvable, except for the near-extreme cases.
    \item For $\iota=10^2$ --- an SNR that may be achieved for very loud current events\footnote{The SNR of the loudest gravitational wave detected to date, GW250114, is 76~\cite{LIGOScientific:2025obp}.} or more commonly by future third-generation ground-based detectors~\cite{Evans:2021gyd,Pieroni:2022bbh,Branchesi:2023} --- the frequency deviation becomes resolvable for $q\gtrsim 0.4$, whereas the damping-time deviation is only detectable close to extremality ($q\gtrsim1.2$).
    \item For an optimistic SNR $\iota=10^3$, which may be accessible for exceptionally nearby sources with third-generation networks~\cite{Evans:2021gyd,Pieroni:2022bbh,Branchesi:2023}, the frequency deviation is distinguishable over most of the $q$ range (except very small $q$), while the damping-time deviation becomes resolvable for $q\gtrsim0.4$.
\end{itemize}

These results indicate that the frequency shifts are generally easier to be detected than the changes in damping time, and that the meaningful discrimination of EMP and RN spectra requires either unusually loud events or next-generation detectors. We emphasize several important caveats: The estimates above are based on the Rayleigh heuristic and leading-order error scalings for single-mode, nonspinning ringdowns in an idealized Gaussian noise; they neglect parameter covariances, multi-mode mixing, merger/remnant modelling systematics, detector calibration uncertainties and non-Gaussian noise transients. We also restrict our attention to the fundamental mode and sample only a limited set of SNRs and scalar charges; population statistics, cosmological redshift and full Bayesian parameter estimation have not been included. A more realistic assessment will therefore require injection studies and a Bayesian inference analysis incorporating multi-mode templates and the full detector noise model.
\section{Conclusion}
\label{sec:conclusion}

The EMP theory can be derived from the EMD theory, where the latter is a four-dimensional low-energy effective theory of string theory. In the EMP spacetime, the phantom scalar field is minimally coupled with Einstein gravity but nonminimally coupled with the Maxwell field. A static and spherically symmetric solution, the so-called charged EB wormhole also describes a regular black hole, providing us a frame to study wormholes or black holes under this spacetime background. 

We investigate the QNMs of the EMP spacetime by studying the linear perturbation of the metric following Chandrasekhar's procedure. Because of the spherical symmetry of the EMP spacetime, the radial parts of the perturbed fields can be decomposed from the angular parts, and the latter can be expanded by the spherical harmonics. Moreover, the perturbation can be classified into the axial (odd-parity) sector and polar (even-parity) sector based on their dependence on the angular coordinate $\varphi$. Note that the phantom scalar field is of spin-0 (even-parity) and therefore does not enter the axial sector. As a consequence, the axial sector contains only the odd-parity metric perturbations and the odd-parity components of the electromagnetic perturbation, forming a self-contained and coupled gravito-electromagnetic system. In the polar sector, the additional physical degrees of freedom emerge, leading to a more complex and physically significant gravito-electromagnetic-phantom scalar coupled system. In order to study the stability of the spacetime under phantom scalar field perturbations only, where the phantom scalar field perturbations are a primary source of potential instability, we adopt a controlled approximation in the polar sector. We next derive the corresponding master equations and rewrite the resulting Schr{\"o}dinger-like equations in a compact matrix form, see Eq.~(\ref{master_eq}). We notice that both the Lagrangian and metric of the EMP theory reduce to general relativity when the phantom scalar hair vanishes and the coupling parameters are chosen properly, so do the master equations and QNFs.

Because of the coupling between the Maxwell field and the phantom scalar field, the effective potential $V_{11}$ contains the coupling function $Z$ in the denominator of  $V_{11}$, see Eq.~\eqref{V GEM}, causing $V_{11}$ to diverge at the zeros of $Z$. We show that these divergence points lie strictly inside the Cauchy horizon in the black hole scenario, whereas they are exposed to the exterior region in the wormhole case.  Moreover, the effective potential contains a negative region when the EMP wormhole is perturbed by a phantom scalar field in the case of $l=0$, leading to a growing time profile obtained via the finite difference method, see Fig.~\ref{fig:wormhole}. These features not only manifest the instability of EMP wormholes but also, by contrast, underscore the perturbative stability of EMP black holes.

We extract the fundamental QNFs from finite difference time profiles using the Prony method and also compute them via the matrix-valued direct integration method. Specifically, We choose $\gamma_1=1/4$ and $\gamma_2=2q$, so that the Type I black hole mimics the RN black hole as the scalar charge $q\to 0^+$. Table \ref{tab:q} tabulates the values of the fundamental QNFs under the phantom scalar field perturbation, the electromagnetic field perturbation and the gravitational field perturbation, respectively, for different values of the scalar charge $q$ with a fixed specific electric charge $\tilde{Q}_\mathrm{e}=2/5$, and Fig.~\ref{fig:QNF_q} provides the corresponding graphical depiction. The results obtained from the Prony method and the matrix-valued direct integration method agree well with each other, which strengthens the validity of our results. As expected, the QNFs are consistent with those of the RN black hole when $q=0.001$.

By introducing the generalized specific charge $\eta$ and the mixing angle $\vartheta$, we are able to study the effect of relative contributions of the scalar and electric charges on the QNMs of Type I black holes, see Figs.~\ref{fig:QNF_PS}--\ref{fig:QNF_Grav} and Tables \ref{tab:QNF_PS}--\ref{tab:QNF_Grav}. We find that the real frequencies and damping rates behave similarly in the phantom scalar perturbation and the gravitational field perturbation, see Figs.~\ref{fig:QNF_PS} and \ref{fig:QNF_Grav} and Tables~\ref{tab:QNF_PS} and \ref{tab:QNF_Grav} for the details, while they show a different behavior only for a small $\vartheta$ and a low $\eta$ in the electromagnetic field perturbation, see Fig.~\ref{fig:QNF_EM} and  Table~\ref{tab:QNF_EM} for the details.

Our investigation into the observability of QNM deviations yields a preliminary but encouraging assessment. Although the spectral shifts between Type I and RN black holes are theoretically significant, they are small: The frequency shifts are more readily measurable than changes in the damping time (see Fig.~\ref{fig:observability}). Given the current gravitational-wave catalog, the robust constraints on the EMP model appear unlikely; nevertheless, future high-precision ringdown spectroscopy --- particularly with third-generation detectors --- could provide a direct empirical test for phantom scalar hairs, transforming these spectral signatures from theoretical predictions into viable observational phenomena.

We focus currently on the linear perturbations of the EMP spacetime, where the interesting features on wormholes and black holes have been summarized above. We shall leave the polar perturbations in the EMP spacetime and a comprehensive observability study to future work and expect new physical insights.

\section*{Acknowledgments}
This work was supported in part by the National Natural Science Foundation of China under Grant No. 12175108.






\end{document}